# Quantitative determination of the level of cooperation in the presence of punishment in three public good experiments

D. Darcet[1] and D. Sornette[2]

[1]Insight Research LLC, Los Angeles, California, USA

[2]Department of Management, Technology and Economics

ETH Zurich, CH-8032 Zurich, Switzerland

ddarcet@noos.fr and dsornette@ethz.ch

tel: +41 (0) 44 63 28917

fax: +41 (0) 44 63 21914

**Abstract:** Strong reciprocity is a fundamental human characteristic associated with our extraordinary sociality and cooperation. Laboratory experiments on social dilemma games and many field studies have quantified well-defined levels of cooperation and propensity to punish/reward. The level of cooperation is observed to be strongly dependent on the availability of punishments and/or rewards. Here, we propose an operational approach based on the evolutionary selection of prosocial behaviors to explain the quantitative level of the propensity to punish in three experimental set-ups. A simple cost/benefit analysis at the level of a single agent, who anticipates the action of her fellows, determines an optimal level of altruistic punishment, which explains quantitatively experimental results on a third-party punishment game, the ultimatum game and an altruistic punishment game. We also report numerical simulations of an evolutionary agent-based model of repeated agent interactions with feedback by punishments, which confirms that the propensity to punish is a robust emergent property selected by the evolutionary rules of the model. The cost-benefit reasoning is not to be taken literally but rather to embody the result of the selection pressure of co-evolving agents that have make them converge to their preferences (which can be seen as either hard-wired and/or culturally selected). In this view, the prosocial preference of humans is a collective emergent process, robustly selected by adaptation and selection. Our main contribution is to use evolutionary feedback selection to quantify the value of the prosocial propensity to punish, and test this prediction on three different experimental set-ups.

# 1-Introduction

The potential for cooperation is everywhere in nature, yet evolution seems to rarely take advantage of it. When it does - social insects, multi-cellular organisms, human societies - the results can be spectacularly successful. One of the most striking characteristics of Homo sapiens is our sociality and cooperation (Fehr et al., 2002). Social relationships pervade every aspect of human life and these relationships are probably far more extensive, complex, and diverse within and across societies than those of any other species. Several theories have attempted to explain the puzzle of human cooperation. (1) Evolutionary kin selection (Hamilton, 1964) roots behavior to genetic relatedness using Hamilton's notion of inclusive fitness. (2) Costly signaling theory (Zahavi, 1977; Gintis et al., 2001) and indirect reciprocity (Nowak and Sigmund, 1995; 1998; Leimar and Hammerstein, 2001) base cooperation in large groups on the build-up of the reputation of cooperators. (3) Reciprocal altruism or direct reciprocity (Trivers, 1971; Axelrod, 1984; Nowak et al., 1995) derives cooperation from selfish motives in the presence of long-term repeated interactions.

However, a growing number of experiments (Kagel and Roth, 1995; Bolton et al., 1998; Henrich et al., 2001; Fehr and Fischbacher, 2003) have documented behaviors which seem incompatible with the self-regarding nature of humans involved in these mechanisms (1-3): humans show a predisposition to cooperate in unrepeated interactions with strangers at their own cost. (4) Strong reciprocity (Gintis et al., 2005) proposes that humans exhibit an inclination to both reward others for cooperative norm-abiding behaviors, and to punish others for norm violations (called respectively "altruistic rewarding" (Fehr and Fischbacher, 2003) and "altruistic punishment" (Fehr and Gächter, 2002)). This provides the potential for group selection and culture evolution (Richerson and Boyd, 2005) above the individual level as a mechanism for spontaneous cooperation within human groups, because of the competitive advantage such cooperation affords. However, at the same time, groups face the problems of internal competition between individuals (or subgroups) and of the free-rider, to which altruistic punishments/rewards constitute a possible remedy.

Economics traditionally conceptualizes a world populated by selfish individualistic agents (Homo economicus) behaving fundamentally so as to maximize their own well-being. In contrast, the theory of strong reciprocity posits that humans are still self-centered but also inequity adverse, that is, their utility function includes terms sensitive to unfairness. But where does this sense of altruistic reward/punishment originate? Alternatively, Bolton and Ockenfels (1999) negate the relevance of altruism; they propose instead that people are motivated by both their pecuniary payoff and their relative payoff standing. People thus have to manage the trade-off between the social reference point and achieving personal gain.

A growing number of researchers are turning to evolutionary biology and biological anthropology in search of the foundations for prosocial preferences. The common thread of an evolutionary explanation of prosocial behaviors is to view strong reciprocity as resulting from natural selection operating on statistically reliable regularities in the environment. The recent discovery of the existence of spontaneous altruism by Chimpanzees and young children adds another evidence for the evolutionary origin of altruism (Warneken et al., 2007). Several evolutionary approaches have been developed to explain prosociality, based on the evolutionary selection of humans interacting in groups and subjected to feedbacks within and between groups resulting in particular from rewards and punishments. For short, we refer to this general class of explanations as "evolutionary feedback selection" theory. A good discussion is provided by the special issue of the journal of Economic Behavior & Organization, organized around the synthesis paper of Henrich (2004). But several issues remain unresolved, such as explaining (a) the level of prosociality (altruism and altruistic punishment) observed in human beings, (b) the patterns of variations of these behaviors across different behavioral domains and social groups, (c) the place of "group selection", and (d) the interactions between cultural and genetic transmissions.

Here, we propose an operational approach based on the evolutionary feedback selection theory to explain quantitatively several experimental results on prosocial behavior: a third-party punishment game, the ultimatum game and an altruistic punishment game. We also report numerical simulations of a simple evolutionary agent-based model of repeated agent interactions with feedbacks, which supports our proposal that evolutionary feedback selection has tuned the level of human propensity to punish to an optimal level in a precise cost-benefit sense for each individual: the trait of providing certain levels of feedbacks (via punishment for instance) has been selected for and has co-evolved with an increasing individual fitness, which is mediated by the collective output of the group. In this sense, the propensity to punish can be seen as an emergent property (Anderson, 1972; Goldenfeld and Kadanoff, 1999).

Our main result is not that altruistic punishments and rewards can emerge from evolution, which is already well-known. Instead, we propose a quantification of the level of cooperation (via punishment) exhibited by humans, which is based on an effective self-centered cost-benefit optimization. We hypothesize that evolution has selected the level of altruistic punishments and rewards in humans which corresponds to an optimal cost-benefit analysis performed on the part of the each agent: it is as if, on average, each agent was doing her optimization calculation to select her level of cooperation and punishment. However, this is only "as if" because, as our agent-based model shows, it is the selection pressure of co-evolving agents that make them converge to their preferences (which can be seen as either hard-wired and/or culturally selected). We do not claim that people actually perform these rational cost-benefit calculations. Rather, humans' individual preferences can be seen as resulting from a collective global organization at the level of the groups, rather than an idiosyncratic optimization. In this view, the prosocial preference of humans is a collective emergent process, robustly selected by adaptation and selection.

The main hypotheses of our approach is that human beings maximise self-interest under the assumption that everyone has seemingly a self-sacrificing tendency to punish selfish persons and that this is a common knowledge. More precisely, people are aware that they can profit from cooperation, that they can gain even more if they are among a minority of non-cooperators with a majority of cooperators, but they also know that they may be punished if they behave unfairly, and they possess a drive to punish other agents who are unfair from their perspective. We propose that these elements have part of the inputs and brain processing abilities that human beings have developed over many generations of cooperation/defection experiences. Our main contribution is to use evolutionary feedback selection theory to quantify the value of the prosocial propensity to punish, and test this prediction on three different experimental set-ups. The underlying basis for our cost-benefit analysis is close to the proposition of Houser et al. (2004) in their discussion of Henrich (2004)'s paper: "An alternative, perhaps simpler, explanation for cooperation is that people have a propensity to reciprocate because it is in their self-interest to do so. Nature has sorted this out over the last 2–3 million years or more, and, in small interdependent social groups, provided humans with the ability to delay immediate gratification in the pursuit of greater mutual gains." Possajennikov (2004) emphasizes similarly that "in the models of… punishment, playing a cooperative action is not altruism, but rather selfish behavior, hoping for future benefits or avoiding costs.'' He adds: "altruists are behaviorally indistinguishable from egoists who socially learn to play the action because of fear of punishment or normative conformity."

Consider for instance Fehr and Gächter (2002; 2003)'s experiments on altruistic punishments, where the adjective "altruistic" refers to a costly behavior with no personal material gain. They show that cooperation flourishes if altruistic punishment is an option, and breaks down if it is ruled out. Fehr and Gächter (2000) have previously shown that spontaneous and uncoordinated punishment activities give rise to heavy punishment of free-riders, that punishment constitutes a credible threat for potential free riders and causes a large increase in cooperation levels. Fehr and Gächter (2002; 2003) suggest that altruistic behavior in simple dilemma games may, in fact, obey rational choice theory, if people indeed pursue altruistic aims, that is, if their utility function includes terms sensitive to unfairness (Fehr and Schmidt, 1999).

Instead of postulating the co-existence of altruistic and selfish aims at the individual level (Fehr and Schmidt, 1999; Fehr and Gächter, 2000), or of conflicting motivations between pecuniary payoff and relative payoff standing (Bolton and Ockenfels, 1999), we propose to use the well-known theory of evolutionary feedback selection to realize that humans are following a consistent approach towards survival and reproduction. Human survival and reproduction, like that of any other species, is the result of an evolutionary process. As social animals, it is likely that human evolution has selected instinctive and emotional behaviors (Damasio, 1994; Bonanno, 2001; Bechara et al., 2000), which promote better survival and reproduction abilities through collective actions. Evolutionary feedback selection can be embedded in bounded rational choice theory, and relates to evolutionary psychology (Ghiselin, 1973; Barkow et al., 1992; Tooby and Cosmides, 1996), whose rationality can only be ultimately understood at the group level, through a kind of "renormalization" by the emergence mechanism (Anderson, 1972; Goldenfeld and Kadanoff, 1999). We interpret evolutionary feedback selection as meaning that feedback levels through rewards and punishments have co-evolved with emotions to ensure high probable survival as well as large collective gains (reproduction success). The potential gain of altruistic reward and the risk of altruistic punishment, that anyone can perceive when acting fairly and unfairly, serve as constraints to one's preference (e.g. quantified by a utility function), promoting better consistency between selfish aims and group efficiency. It becomes in anyone's interest to act fairly under the feedbacks provided by the anticipation of other group members' altruistic reward and the threat of other group members' altruistic punishment.

We first propose a general quantitative formulation of these concepts in section 2, which is then applied to the third-party punishment game (section 3), to the ultimatum game (section 4) and to altruistic punishment games (section 5). In the first two games, unfairness appears in the payoffs while the contributions of the agents are irrelevant. In the third game, the payoffs are equal but inequity appears via possible different contributions to the common project. Section 6 presents a simple agent-based model with repeated interactions with punishment. Section 7 briefly addresses the asymmetry between the two feedback processes: punishment versus reward. Section 8 concludes.

**2-General formulation**

Let us consider n agents in a social dilemma situation, in which voluntary contributions are needed to obtain some shared end-result, and where the individual rational choice is to free-ride. We denote by $c_i$ i=1, …, n, the contributions of each agent to the common project. Then, the shared end-result is quantified by n payoff functions $P_i(c_1, c_2, …, c_i, …, c_n)$, which are possibly distinct. After the one-period play, the total wealth of agent i is therefore $P_i(c_1, c_2, …, c_i, …, c_n) – c_i$. The conditions for a social dilemma to hold is that

$P_i(0, 0, …, 0, c_i, 0, …, 0) < c_i$ , (1a)
$P_i(c_1, c_2, …, c_{i-1}, 0, c_{i+1}, …, c_n) > P_i(c_1, c_2, …, c_i, …, c_n) - c_i$    for all i's. (1b)
$P_i(c, c, …, c, c, c_1, …, c) > c$ . (1c)

Condition (1a) writes that the project does not remunerate sufficiently the individual contributions and thus discourage agents to contribute independently of the actions of the other agents. Condition (1b), which should hold for arbitrary n-plets ($c_1$, $c_2$, ..., $c_i$, ..., $c_n$), implies that the rational choice is to free-ride. Condition (1c) expresses the fact that it pays to contribute if the others also contribute. It is the competition between (1a) and (1b) on the one hand and (1c) on the other hands that creates the social dilemma.

It is well known that the introduction of punishment opportunities after every public goods interaction creates high incentives to cooperate and contribute to the public good. However, because punishment is costly, social behavior may also get undermined by non-punishing strategies and subsequently cooperation breaks down. This is the second-order cooperation problem in which the rational behavior is not to punish. In order to get persistent cooperative behavior, additional mechanisms are required. One such mechanism is reputation, which is built up during past encounters. Here, we emphasize another mechanism, associated with the collective co-evolution of cooperation and feedback by punishment in a population of agents under repeated interactions. In other words, in the mechanism studied by the theory of evolutionary feedback selection, cooperation and punishment are emergent characteristics within an evolving population.

According to evolutionary feedback selection, agents have evolved the possibility to exert a feedback on other agents through rewards and/or punishments. We suggest that the level of reward/punishment is determined from the evidence that people assign and maintain a self-perspective both in action and social interactions (Vogeley et al., 2001; Vogeley and Funk, 2003) within their 'theory of mind' (Flechter et al., 1995; Gallagher and Fritch, 2003), which is the capacity to attribute opinions, perceptions or attitudes to others. The first-person-perspective suggests that an agent will measure the unfairness of the allocation by comparing pair wise contributions and pair wise payoffs. We first discuss the mechanism of feedback by punishment and refer to section 7 for the case where the feedback is by reward. The feedback by punishment should take into account the following situations:

(i) The contributions are different but the payoffs are identical, a situation often encountered in global resource sharing, such as for instance in public good interactions and in the tragedy of the commons. The level of punishment exerted by an agent k on an agent i observing the contributions $c_i$ and $c_j$ of two agents i and j (k can also be j herself) should be an increasing function of $|c_i - c_j|$, with the punishment applied to the smallest contributor. Here, the motivation is to punish the free rider.

(ii) The contributions are identical but the payoffs are different, perhaps due to random factors or to some structural asymmetry. This situation is characteristic of the ultimatum and dictator games for instance. The level of punishment exerted by an agent k (on agent i) observing the payoffs $P_i$ and $P_j$ of the two agents i and j should be an increasing function of $|P_i - P_j|$ with the punishment applied to the greatest payoff. Here, the rational is to punish the largest unjustified/unfair endowment.

(iii) A mixture of (i) and (ii) in which both contributions and payoffs are different. The arguably simplest functional form accounting for these situations is

$$Pu_{k \to i} = \sum_{[j=1 \text{ to } n]} \kappa_{kji}^{c}(c_j - c_i) + \sum_{[j=1 \text{ to } n]} \kappa_{kji}^{P}(P_i - P_j). \qquad (2)$$

$Pu_{k \to i}$ is the punishment of agent k on agent i, based on the observation of both the contributions and payoffs of agent i and of all the other agents j's. It is estimated as a weighted sum of functions of the observed differences of the contributions and payoffs of all agents j and the agent i. We assume that $\kappa_{kji}^{c}(c_j - c_i)$ and $\kappa_{kji}^{P}(P_i - P_j)$ are non-decreasing functions of their arguments. One could also consider coupled terms combining ($c_j - c_i$) and ($P_i$

– $P_j$). In this first approach, we consider functions which are linear by part: specifically, we expand the function as follows: $\kappa_{kji}^c(c_j - c_i) = \kappa_{kji}^c * (c_j - c_i)$ for $c_j - c_i > 0$ and 0 otherwise. Here, the symbol $*$ stresses that the notation $\kappa_{kji}^c(c_j - c_i)$ for the function is replaced by a proportionality factor $\kappa_{kji}^c$ times the discrepancy $(c_j - c_i)$ of the contributions. Similarly, we replace $\kappa_{kji}^P(P_i - P_j)$ by $\kappa_{kji}^P * (P_i - P_j)$ for $P_i - P_j > 0$ and 0 otherwise. In this linearization scheme, coupled terms combining $(c_j - c_i)$ and $(P_i - P_j)$ are higher order in a general functional expansion and are not considered. The weight factors $\kappa_{kji}^c$ and $\kappa_{kji}^P$ (which we refer to as "propensities to reward/punish") can be heterogeneous to reflect prior knowledge such as reputation or to take into account that not all agents have the ability to punish. Note the change of order of the indices in the two sums in (2), reflecting the following property: the punition of agent k on agent i is activated if the contribution $c_i$ tends to be smaller than the other agents $c_j$'s and/or when its payoff $P_i$ tends to be larger than those of the other agents. In the sequel, when the situation allows, we use the simplification that the $\kappa_{kji}^c$ (resp. $\kappa_{kji}^P$) are the same $\kappa^c$ (resp. $\kappa^P$) for all agents k and all pairs (i, j). In this special case, expression (2) simplifies into

$$Pu_{k \to i} = \kappa^c (n-1) (<c_j>_i - c_i) + \kappa^P (n-1) (P_i - <P_j>_i), \qquad (3)$$

where $<c_j>_i$ (resp. $<P_j>_i$) is the average contribution (resp. payoff) of the n-1 agents other than agent i.

We hold that people are aware that they can profit from cooperation, that they can gain even more if they are among a minority of non-cooperators with a majority of cooperators, but they also know that they may be punished if they behave unfairly, and they possess a drive to punish other agents who are unfair from their perspective. We propose that these elements have part of the inputs and brain processing abilities that human beings have developed over many generations of cooperation/defection experiences. In this context, the level of cooperation of a given agent i is assumed to be determined by the following optimization problem that a given agent i must solve to determine her contribution $c_i$

$$\text{Max}_{c_i} \ E[ P_i(c_1, c_2, ..., c_i, ..., c_N) - \sum_{[k=1 \text{ to } n]} Pu_{i \to k} - r_p \sum_{[j=1 \text{ to } n]} Pu_{j \to i} ] \qquad (4)$$

In expression (4), E[ . ] denotes the expectation of agent i and $r_p$ is the so-called reward/punishment efficiency parameter (see below for its precise meaning in the three treated examples). The optimization problem for agent i consists in determining her optimal contribution $c_i$ based on her expectation of the contributions of the other agents, which determines her expectation of the punishment she may have to endure from other agents and the punishment she may be inclined to impose on the other agents.

This formulation (4) is reminiscent of the approach in terms of fairness utility functions, which take it as (exogenously) given that people have preferences for fair contributions (Henrich et al., 2004). The fact that cooperation is stable given such fairness preferences then emerges naturally, since contributors will anticipate punishment. Our approach differs by (i) seeing fairness as resulting from the propensity to punish in the presence of unfairness (rather than being an innate contribution to one's utility) which allows us to (ii) determine quantitatively the level of punishment. In particular, the main target of our approach is to determine endogenously the quantitative value of the propensity to punish by a condition of maximum selfish gain in an evolutionary stable collective equilibrium point. The next critical step, which is to explain how such punishment behavior could evolve in the first place, is illustrated in section 6 within an agent-based simulation.

We use the theory of evolutionary feedback selection to determine the coefficients $\kappa_{kji}^c$ and $\kappa_{kji}^P$ so as to ensure the cooperation of the members of the group (i.e. the resolution of the free-rider social dilemma), and to increase the average pay-offs for each member of the group compared with the Nash equilibrium without social cooperation. According to the evolutionary feedback selection theory, the evolution of the propensity to punish has occurred endogenously and self-consistently via the existence of the feedback mechanism (2) which has been active over many generations. We propose that the propensity to punish has adjusted itself so that the marginal expected gain of any agent becomes zero, yielding a Pareto efficient dynamical fixed point. In the examples treated below, we implement this program and show how it allows us to determine the Pareto optimal propensity to punish, which is compared quantitatively with experimental results.

**3-Analysis of a third-party punishment game**

Fehr and Fischbacher (2003) report experiments on a variant of the ultimatum game, in which the possibility of punishing is not given to the recipient, but to a third party not involved in the allocator-recipient collective action. The existence of punishment in this experiment materializes what Adam Smith called "sympathy" in his Theory of Moral Sentiments, defined in his words as "our fellow-feeling with any passion whatever".

In the experiment reported in (Fehr and Fischbacher, 2003), an allocator is endowed with 100 MUs (Monetary Units) and is given the free option to donate part of her endowment to a recipient, who had received nothing at the beginning of the game. A third party, who receives 50 MUs, can spend money to punish the allocator when she is informed of the money transfer from the allocator to the recipient. Every MU spent on punishment by the third party reduces the allocator's wealth by $r_p=3$ MUs. The coefficient $r_p$ can be thought of as the "punishment efficiency." The expected rational action is for the third party not to punish whatever the transfer from the allocator to the recipient, since she can only lose if she punishes. This expectation is contradicted by experimental observations (Fehr and Fischbacher, 2003).

In the notations of section 2, we have n=3. Let us refer to the allocator with the index 1, the recipient with the index 2 and the third party with the index 3. Then, by the structure of the game, all contributions are identical (or irrelevant) and only the payoffs need to be taken into account. In addition, by construction, all $\kappa_{kji}^P$ are zero except $\kappa_{321}^P$ which will be called 'k'. The payoff functions are $P_1 = 100 - m_a$, $P_2 = m_a$ and $P_3=0$.

In the framework of our approach exploiting the concept of evolutionary feedback selection, the expected gain $G_a$ of the allocator, who transfers $m_a$ MUs to the recipient, is:

$$E[G_a \mid m_a] = 100 - m_a - r_p E[Pu_{\text{third party} \to a}] \qquad (5)$$

It is the sum of her remaining endowment $100-m_a$ after the transfer of $m_a$ MUs to the recipient, decreased by the specter of the punishment by the third party. The punishment term $E[Pu_{\text{third party} \to a}]$ must express the fact that the third party will measure the unfairness of the allocation by comparing the two payoffs. The simplest assumption, justified as a Taylor expansion of an arbitrary non-singular function limited to the first-order, is that the punishment $Pu_{\text{third party} \to a}$ of the third party onto the allocator is proportional to the mismatch or deviation from fairness $(100-m_a) - m_a$ of their gains:

$$E[Pu_{\text{third party} \to a}] = k [(100-m_a) - m_a] \quad \text{if } m_a \leq 50 \qquad (6)$$
$$E[Pu_{\text{third party} \to a}] = 0 \quad \text{if } m_a > 50$$

where k is a proportionality factor that the theory must determine self-consistently. Note that expression (6) is nothing but (2) with all $c_i$'s being equal and all $\kappa_{kji}^P$ equal to zero except $\kappa_{321}^P=k$. In (6), we take a threshold $m_a=50$ corresponding to an average preference for 50%-50% splits. This assumption is not crucial and can be relaxed: taking for instance a 60%-40% split preference actually improves the fit between the prediction of the theory and the realized punishments shown in figure 1. Our formalism (2) can account for different cultures or contexts in which the preferred split is not 50%-50% (Camerer, 2003), by interpreting the asymmetry between allocator and recipient in terms of distinct contributions $c_1$ and $c_2$. In this first approach, we do not explore further this interesting line of investigation. We refer to the rule (5) as the "linear punishment response function." The rule is assumed to be known (endogenized consciously or instinctively) by both the allocator and the third party, reflecting a common evolution of feedback rewards and punishments over many generations.

According to evolution feedback selection, one can see the level of punishment of a human being playing this game in modern times as resulting from hard-wired emotional decision modules selected by evolution over many generations. In the experiments reported in (Fehr and Fischbacher, 2003), third parties punish allocators at their own cost without any monetary gain, despite the fact that they will never see each other in the future. We interpret evolutionary feedback selection as saying that the interplay between selfish optimization and feedback has tuned the level of punishment so that the third party can empathize with the allocator, and determine that the latter will take into account her gains as well as her potential punishment, in such a way to maximize her expected gain.

Quantitatively, the theory proceeds as follows: we calculate the marginal expected gain of the allocator who hesitates between transferring $m_a$ or $m_a+1$ MUs and use the evolutionary self-centered optimization to determine the punishment coefficient k. The marginal expected gain of the allocator reads

$$E[G_a | m_a+1] - E[G_a | m_a] = -1 + 2 k r_p \quad \text{if } m_a \leq 50 \quad (7a)$$
$$E[G_a | m_a+1] - E[G_a | m_a] = -1 \quad \text{if } m_a > 50 \quad (7b)$$

Suppose that an initial guess is $m_a>50$. Then, (7b) tells the agent that she is better off by reducing her transfer. When she tests her expected gain with a transfer of 50 or less, she has two alternatives: (i) either she believes that the propensity k to punish of the third-party is smaller than $1/2r_p$, which implies that her optimal strategy is to transfer the minimum $m_a \Rightarrow 0$, corresponding to the non-cooperating fixed points. This is a possibility. Another possibility is that she believes that $k>1/2r_p$, in which case, she should transfer 50. In this case, the cost to punish for the third-party is zero but any deviation $m_a<50$ from the fair allocation is expected to cost the third-party proportionally to this deviation and to k. We hypothesize that evolution has selected the k value which leads to the most efficient allocation of resources, that is, the smallest cost to the punisher, given the possible deviation from the fair allocation $m_a=50$. In other words, the cheapest altruistic punishment to ensure cooperation is reached when

$$k \rightarrow 1/2r_p \quad (8)$$

The theoretical punishment level then reads:

$$E[Pu_{\text{third party} \rightarrow a}] = (1/2r_p) [(100-m_a) - m_a] \quad \text{if } m_a \leq 50 \quad (9)$$
$$E[Pu_{\text{third party} \rightarrow a}] = 0 \quad \text{if } m_a > 50$$

Figure 1 compares this prediction (9) with the punishment beliefs by recipients and with the actual punishment by third parties, as reported by Fehr and Fischerbach (2003). Given the fact that there are no adjustable parameter (since $r_p$ is fixed to the value used in (Fehr and Fischbacher, 2003)) and that we use the approximation of a linear punishment response function (6), the agreement is good.

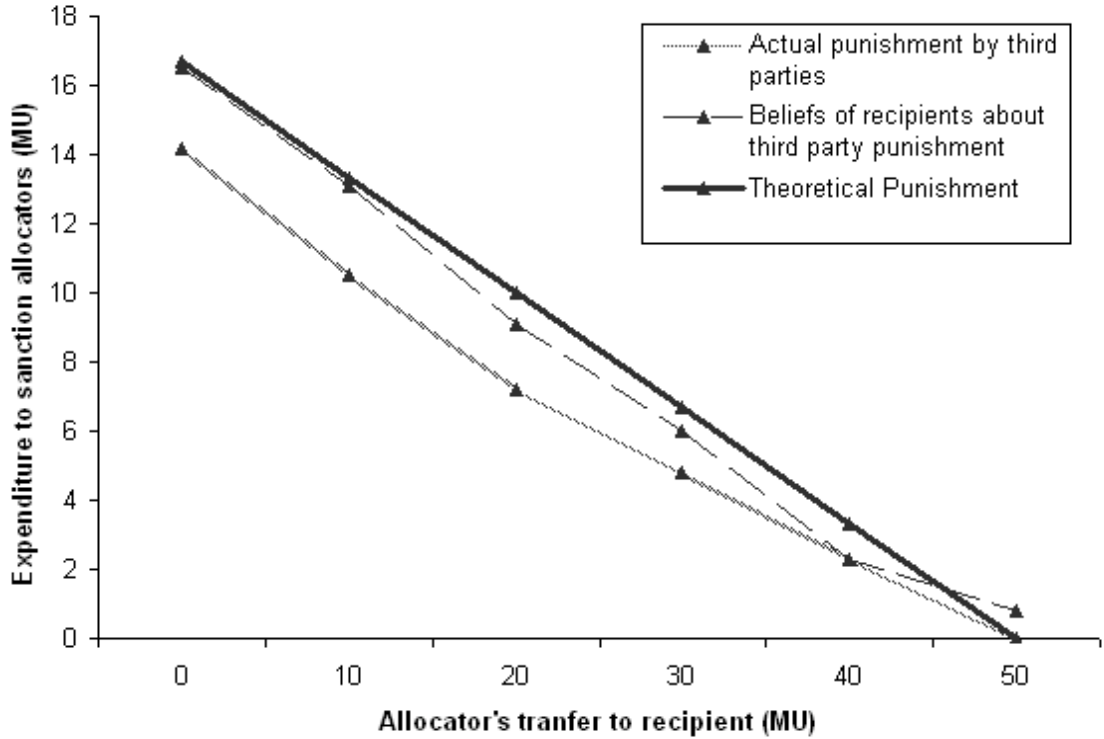

**Fig.1:** Mean expenditure by the punishing third party (in MUs) as a function of the transfer by the allocator to the recipient (in MUs) obtained in the experiments reported in (Fehr and Fischbacher, 2003). The theoretical punishment represented by the thick line is given by expression (9) with the value $r_p=3$ of the punishment efficiency used in (Fehr and Fischbacher, 2003).

**4-The ultimatum game**

In the ultimatum game, two subjects have to agree on the division of a fixed sum of money, for example 100 MUs. The proposer makes exactly one proposal $0 \leq m_a \leq 100$ of how to divide the money. Then, the responder can accept or reject the proposed division (100-$m_a$ to the proposer and $m_a$ to the responder). In the case of rejection, both receive nothing, whereas in the case of acceptance, the proposal is implemented. The rational solution is for the responder to accept any positive transfer, and thus for the proposer to propose the minimum possible non-zero amount. A robust result in this experiment across many cultures is that responders reject the division with a high probability if $m_a$ is too low (Henrich et al., 2004; Oosterbeek et al, 2004). The ultimatum game belongs to the class of games with altruistic punishment, in the sense that the responder often sacrifices his share $m_a$ as his cost for punishing the proposer with no personal gain.

Analyzed from the viewpoint of the proposed cost-benefit analysis, the particularity of this game is that the punition leverage coefficient $r_p$ previously defined in (4) and (5) now depends on the contribution $m_a$ itself. Indeed, using expression (6) and the fact that punishment is equivalent to rejection by the responder which amounts to losing the entire sum 100-$m_a$ for the proposer, this leads to define a punishment efficiency $r_p = (100-m_a)/m_a$, equal to the ratio of the loss incurred by the proposer (punished) to the loss incurred by the responder (punisher). This game differs from the previous third party game and from the altruistic punishment game discussed below by the fact that its punishment efficiency $r_p$ is not fixed but is determined by the level of cooperation quantified by $m_a$. A small offer $m_a$ from the proposer to the responder not only implies a poor cooperation but also a strong punishment efficiency, both factors favoring a strong potential for punishment. In the notations of section 2, using the index 1 for the proposer and the index 2 for the responder (n=2), we have all contributions being a priori equal (or irrelevant), only payoffs enter in the punishment function (2) with all $\kappa_{kji}^P$ being zero except $\kappa_{221}^P$, which will be called `k'. The payoff functions are $P_1=100-m_a$ and $P_2=m_a$.

Proceeding as for the preceding example, we write the expected gain $E[G_p | m_a]$ of the proposer, given a proposal $m_a$:

$$E[G_p | m_a] = 100-m_a - k\, r_p\, (100-2m_a) = 100-m_a - k(100-m_a)(100-2m_a)/m_a \qquad (10)$$

This expression explicitly assumes that the expected fair offer is 100/2. Deviations from this value, perhaps due to different cultures and/or to particular circumstances, are straightforward to take into account. The marginal gain of the proposer who hesitates between proposing $m_a$ and $m_a+\eta$, where $\eta$ is a small increment, is

$$\eta\, dE[G_p | m_a]/dm_a = \eta\, [-1 + k\, (10^4/m_a^2 - 2)] \qquad (11)$$

Taking the expectation of this expression, with respect to the distribution of propositions $m_a$, and equating to zero, allows us to determine the coefficient k of punishment, which results from the evolution of selfish maximizers in the presence of punishment feedbacks:

$$k = E[1/(10^4/m_a^2 - 2)] \qquad (12)$$

The probability $p_r(m_a)$ that the responder rejects the proposition $m_a$ can be obtained from the fact that the expected gain of the proposer can be written as the product of her endowment multiplied by the probability $1-p_r$ that the offer is accepted: $E[G_p | m_a] = (100-m_a)(1-p_r)$. This yields

$$p_r = k\, (100-2m_a)/m_a, \qquad \text{where k is given by (12)} \qquad (13)$$

According to the theory of evolutionary feedback selection, the level of altruistic punishment in the ultimatum game should be highly sensitive to cultural and/or economic differences, measured by the $m_a$ distribution. The factor k should have co-evolved with social norms represented by the distribution of "adequate" allocations by the proposer. Human groups exhibiting a low propensity to cooperate in the ultimatum game should as well select low rejection rates. Such sensitivity is indeed observed in laboratory experiments, see for instance the comparison between Indian and French participants (Boarini et al., 2004).

Figure 2 and figure 3 compare the theoretical and experimental rejection rates in ultimatum games involving French and Indian participants (Boarini et al., 2004). Experimental observations on French participants are well anticipated by the theory while one can observe significant discrepancies between theory and data for Indian participants. Interestingly, this discrepancy comes with another important difference between French and Indian behaviors, as reported by Boarini, Laslier and Robin (2004). Indeed, while French's offers remain constant over repetition of the game, suggesting that the appropriate level of punishment leads to a stable collective feedback process, Indians' offers continuously decrease over time as the games are repeated. This suggest that the observed level of rejection of Indians is not a true characteristic associated with a stable cooperating behavior but rather a transient learning phase. The conclusion that one can draw from our analysis is that Indian rejections rates are not optimized to the experiment performed in (Boarini et al., 2004), leading to cooperation rapidly vanishing. This may be due to a wealth effect (obviously, people at the marginal level of subsistence will not turn down an offer, however small it is; they do not have the luxury to punish). We should state carefully that we do not say that evolution has led Indian people to be less cooperative than French people (!); we only remark that the conditions of the experiment with the Indian people are not comparable with that with the French People and one should control for other economic parameters in order to perform a meaningful comparison. Note also that some differences in the rejection rates of participants in different countries (such as Israel, Japan, USA, Yugoslavia) have been documented, which can be explained by different expectations about what constitutes an acceptable offer (Roth et al., 1991), a feature that can be easily incorporated in our formalism.

We note also that more complex situations arise under repetitions of the ultimatum game, which lead to the interplay with and cross-over to bargaining effects (Binmore et al., 1985). However, experiments on repeating bargaining games confirm that bargainers incorporate fairness concerns in their propositions (Ochs and Roth, 1989), a characteristic which can be readily explained by our approach and does not require assumptions on the utility function of agents.

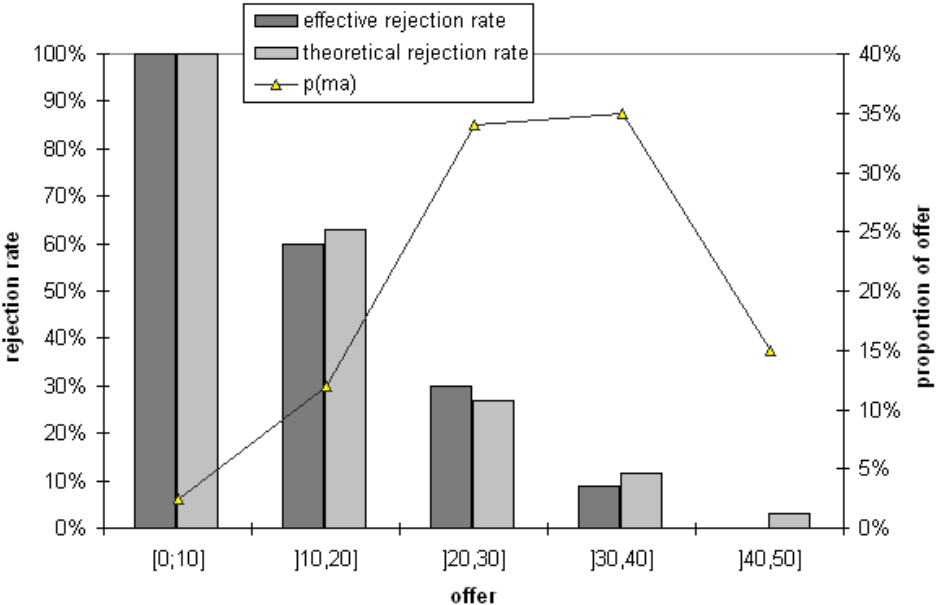

**Fig.2:** Experimental rejection rates in an ultimatum game involving French participants performed by Boarini, Laslier and Robin (2004) compared with theoretical rejection rates. The theoretical rejection rate is calculated using expressions (13) with (12), where the expectation in (12) is performed over the observed distribution $p(m_a)$ reported in (Boarini et al., 2004), which is shown as the continuous line joining the triangles. There is thus no adjustable parameter in the theory. Note that the cooperation reported in (Boarini et al., 2004) is stable over time.

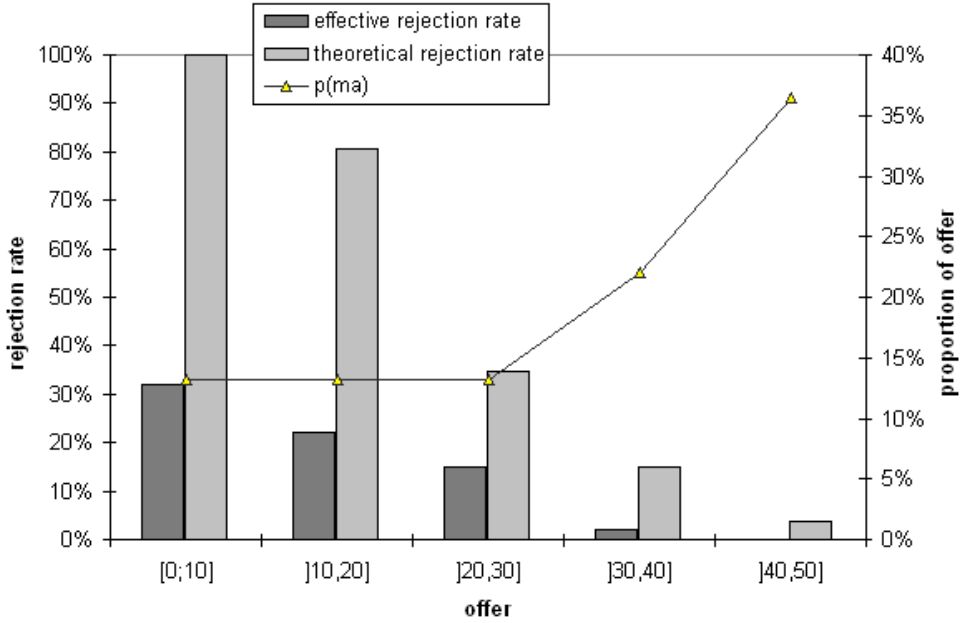

**Fig.3:** Experimental rejection rates in an ultimatum game involving Indian participants, performed by Boarini, Laslier and Robin (2004), compared with theoretical rejection rates obtained with the procedure explained in the caption of figure 2. Indians' offers are observed to decrease with repetition of the game leading to rapidly vanishing cooperation, showing the absence of a stationary state. The conditions for application of the theoretical predictions (12) and (13) are thus not met.

## 5-Analysis of Fehr and Gächter (2002; 2003)'s experiments on altruistic punishments
### 5.1 Quantitative formulation

In the experiments performed in (Fehr and Gächter, 2002), n=4 members of a group receive each an endowment of M=20 monetary units (MUs), and each can contribute to a project between 0 and 20 MUs. One MU invested by the group in a project yields back $r_1$=1.6 MUs. This yield is equally shared among the n participants, which implies that 1 MUs invested by an agent is returned as $r_1/n$=0.4 MUs to her, that is, with a loss of 1- $r_1/n$=0.6 MUs: it is thus always in the material interest of any subject to keep away from the project. However, if all members follow this reasoning and do not contribute to the project, they just earn their initial endowment of 20 MUs while, if they all contribute maximally, they earn each 20 $r_1$ = 32 MUs, showing that this game is an example of a social dilemma. Indeed, a free-rider among n-1 maximally cooperating agents will earn even more, that is, 20+20(n-1)$r_1/n$=44 MUs. Without feedbacks, rational decision theory predicts that cooperation should break down after a few rounds, which is indeed observed experimentally.

However, more interesting behaviors occur when feedback by punishment is introduced. After the above first step, according to the rules implemented in the experiments of (Fehr and Gächter, 2002), each member j of the group can decide to spend $m_{j \to i}$ MUs to punish another agent i (agent j can punish several agents) whom she feels has not contributed enough to the group, where $0 \leq m_{j \to i} \leq m_p$=10. By spending $m_{j \to i}$ MUs, agent j inflicts a cost equal to $r_p m_{j \to i}$ MUs to a punished member i (where $r_p$=3 is the efficiency of the punishment). Each agent is aware of this possibility before she decides the level of her contribution to the group. In the notations of section 2, n=4, all payoffs are identical but the contributions differ. All coefficients $\kappa_{kji}^c$ are a priori non-zero and, in view of the symmetry between all agents, we assume that they are all equal to some common value `k.'

In the experiments reported in (Fehr and Gächter, 2002), group members punish each other, despite the fact that they will never see each other in the future. Again, we use evolutionary feedback selection to postulate that the interplay between selfish optimization and feedback has tuned the level of punishment (quantified by the propensity k to punish) so that an agent, who takes into account her gains as well as both her potential cost to punish and her potential punishment by others, has a zero average expected marginal gain. The experiment shows that defectors with very low or zero investment in the project incur a larger probability of being punished. Our goal is to predict the observed experimental expenditure by punishing group members as a function of the deviation from mean cooperation level of other group members.

As an agent does not know the contributions of the other members when she decides her own contribution, she must form a judgment on the distribution of other members' contributions. Given this judgment, she adjusts her contribution according to some criterion of optimization. Following the same reasoning as in the previous sections, we propose the simple rule that the agent maximizes her expected gain, given her anticipation of the distribution of contributions by the other players and the existence of punishment.

If $m_1, m_2, \ldots, m_{i-1}, m_{i+1}, \ldots, m_n$ are the amounts of MUs invested by members 1, 2…,i-1, i+1,…, n in the project and which are unknown to agent i, the expected gain $E_i[G_i | m_i]$ for agent i in absence of punishment, and conditioned on her contribution $m_i$, is:

$$E_i[G_i | m_i] = -m_i + (r_1/n) E[m_1+m_2+\ldots+m_n] = -m_i(1-r_1/n) + (r_1/n) E[m_1+m_{i-1}+m_{i+1}+\ldots+m_n]$$
$$= -m_i(1-r_1/n) + (n-1)(r_1/n) E_i[m_j] \qquad (14)$$

where $E_i[m_j]$ is the expected contribution of another typical agent $j \neq i$. The last equality expresses the fact that the agent i does not have any prior information on the other members of the group and thus forms identical expectations on their contributions. Thus, from the viewpoint of agent i, all other members of the group are inter-changeable in absence of prior information. This is a prerequisite of the experimental conditions described in (Fehr and Gächter, 2002). Therefore, it is convenient to split agent i's total expected gain into a series of one-to-one gains with respect to each other member. This reads

$$E_i[G_i | m_i] = \sum_{[j=1 \text{ to } n; j \neq i]} E_i[G_{i,j}] \qquad (15)$$

with

$$E_i[G_{i,j} | m_i] = -a\, m_i + (r_1/n) E_i[m_j], \quad \text{with} \quad a = (1-r_1/n)/(n-1)] > 0, \quad \text{for } j \neq i. \qquad (16)$$

Since $r_1/n < 1$, a selfish agent has the incentive to defect ($m_i = 0$) to maximize $E_i[G_{i,j} | m_i]$, but profits from others' cooperation ($m_{j \neq i} > 0$) embodied in $E_i[m_j]$. We assume that agent i uses a (possibly subjective) probability $P_j(p)$ to quantify her belief that the agent j will invest p MUs in the group project. We can thus write

$$E_i[m_j] = \sum_{[p=0 \text{ à } 20]} p\, P_j(p) \qquad (17)$$

In addition to the accounting of her potential gains given by (15) and (16), in order to make her decision on how much to contribute, agent i takes into account (possibly through not fully conscious decision making modulii developed by evolution) both her potential cost to punish and her potential punishment by others. For these two additional contributions, our starting point is, once again, the evidence that people assign and maintain a self-perspective both in action and social interactions (Vogeley et al., 2001; Vogeley and Funk, 2003) within their 'theory of mind' (Flecher et al., 1995; Gallagher and Fritch, 2003). The first-person-perspective suggests that a member j will measure the lack of cooperation of agent i by comparing the contribution $m_i$ of agent i to the group project to her own contribution $m_j$. The simplest assumption, justified as a Taylor expansion of an arbitrary non-singular function limited to the first-order, is that the punishment $Pu_{j \to i}$ of agent j on agent i is proportional to the mismatch $m_j - m_i$ of their contributions, as in expression (2):

$$Pu_{j \to i} = k\,(m_j - m_i), \text{ for } m_j > m_i \ ; \qquad Pu_{j \to i} = 0, \text{ for } m_j \leq m_i \qquad (18)$$

This "linear punishment response function" has the same meaning as expression (2) used for the two previous games. Predicting the value of the positive coefficient k is one of the major goals of the theory as it quantifies the propensity to punish. The total expected gain of agent i in her relation with agent j, which takes into account both possibilities that she punishes agent j or agent j punishes her, conditioned to her contribution $m_i$, then reads

$$E_i[G_{i,j} \mid m_i] = -a\,m_i + (r_1/n)\,E_i[m_j] - r_p\,E_i[Pu_{j \to i}] - E_i[Pu_{i \to j}] \qquad (19)$$

where $r_p$ (=3 in [1]) is the punishment efficiency and

$$E_i[Pu_{i \to j}] = k \sum_{[p=0 \text{ to } m_i]} (m_i - p)\,P_j(p) \ ; \ E_i[Pu_{j \to i}] = k \sum_{[p=m_i \text{ to } 20]} (p - m_i)\,P_j(p) \qquad (20)$$

The term $E_i[Pu_{i \to j}]$ is the expectation of agent i of how much she will punish another typical agent j, given her contribution $m_i$. The term $r_p\,E_i[Pu_{j \to i}]$ is the expectation of how much punishment she will endure from agent j, conditioned on her contribution $m_i$.

The total expected marginal gain of agent i, who hesitates between investing $m_i$ and $m_i + 1$ (with i<20), then reads

$$E_i[G_{i,j} \mid m_i+1] - E_i[G_{i,j} \mid m_i] = -a + k\,[\,r_p\,(1-F_i) - F_i\,] \qquad (21)$$

where
$$F_i = \sum_{[p=0 \text{ to } m_i]} P_j(p) \qquad (22)$$

is the probability perceived by agent i that another agent j will contribute less than or equally to her. Equation (21) expresses the fact that an increment of 1 MU of agent i's contribution gives conflicting terms: (1) an anticipated loss "–a" due to the equal sharing of any investment between team members, (2) an anticipated gain $kr_p(1-F_i)$ due to the decrease of the probability of being punished by agent j (which increases with the probability $1-F_i$ that the other agent contributes more than herself) and (3) an anticipated loss $-kF_i$ due to the increase in the probability of punishing the other agent j (which grows with the probability $F_i$ that the other agent contributes less than herself).

According to our interpretation of the theory of evolutionary feedback selection, the level of feedback provided by punishment, quantified by the factor k, has been optimized globally over many human generations so that the selfish optimization of an arbitrary agent ensures a maximum gain when investing in the group project. The value of k has been selected and transmitted to ensure that cooperative actions are maximally efficient from the self-centered point of view of an agent. As a consequence, humans unconsciously select a level of punishment, which has been fine-tuned by the evolutionary success of cooperation kept under control by the feedback provided by punishments. Similarly, in more general situations, they may select group members they are able to understand and anticipate. The ideal situation is reached when agent i invests exactly the same amount as agent j, because no one incurs any punishment cost. This performance is heavily dependent on the ability of agent i to anticipate the contribution of agent j and reciprocally.

Whether the selection and transmission of the optimal level of feedback has occurred through biological and/or cultural processes can make a significant difference in some situations (Boyd and Richerson, 1985), as for instance cultural transmission can result in transmission of maladaptive traits. Effects reminiscent of this result are observed in our agent-based model in section 6, in which some groups are found to evolve spontaneously towards the non-cooperative Nash equilibrium.

Since all players play a symmetric role, the distribution of contributions $m_i$ describing the heterogeneity of the players' contributions is the same as the distribution $P_j(p)$ of the contributions of the other agents. The fact that the level of punishment k has been selected by evolution and cultural interactions implies that it constitutes a collective characteristic of human cultures. But k is not the sole parameter characterizing the evolved adaptation. The distribution $F_i$ can also be expected to adapt to the punishment efficiency $r_p$, as shown in a simple agent-based model discussed in section 6.

In order to determine k, we consider the following situation. Let us assume that the adaptation is dominated by many stochastic factors, so that the only robust response for k is to maximize $E_i[G_{i,j} | m_i]$, i.e., make (21) vanish after averaging all possible contribution levels $m_i$ weighted by their corresponding probabilities $P_j(m_i)$. Since by construction the random variable $F_i$ is uniformly distributed between 0 and 1, its average with respect to $m_i$ weighted by $P_j(m_i)$ is always equal to ½, irrespective of the arbitrary distribution of $P_j(m_i)$ of the contributions of the agents. One can interpret this result by saying that the average probability for a contribution to be smaller than or equal to the median is exactly ½. In other words, the best guess for the contribution of others is the median (which is however not specified and disappears from the unknowns by the averaging procedure). Then, equating the average of (21) (with $E[F_i]=1/2$) to zero to express the condition of a maximum expected gain yields

$$k = 2a/(r_p-1) = 2(1-r_1/n) / [(n-1)(r_p-1)] \qquad (23)$$

For $n=4$, $r_p=3$, and $r_1=1.6$, we obtain $k=0.2$. Accordingly, humans will punish (i.e., $k>0$) typically in situations of social dilemma ($r_1/n<1$) and when punishment is effective ($r_p>1$). It is important to realize that this result (23) amounts to assuming that the distribution of contribution is quenched, i.e., does not respond to the punishment efficiency $r_p$.

The simulations of the agent-based model discussed below as well as experiments (Bernhard et al., 2006; Anderson and Putterman, 2006) suggest that the distribution $P(p)$ is effectively distorted for large values of $r_p$ so that $F(<m_i>)$ tends to increase with $r_p$ and the factor k remains almost constant as a function of $r_p$. This phenomenon is neglected by the averaging of $F_i$ but, for the value $r_p=3$ of the experiment of (Fehr and Gächter, 2002), the factor k selected in the agent-based model is undistinguishable from the value $k=a$ predicted by equation (23).

The punishment given by agent j, when she realizes that agent i has invested unfairly, is given by expression (18) with (23). The total punishment exerted by the group of n-1 agents towards agent i sums up the individual punishments from each of the n-1 other agents participating in the collective action. The total expenditure by punishing members towards agent i reads:

$$Pu_{group \to i} = \sum_{[j, ; j \neq i, m_j > m_i]} Pu_{j \to I}$$
$$Pu_{group \to i} = [2(1-r_1/n)/(r_p-1)](<m_{j; j\neq i, , m_j>m_i}> - m_i) = 0.6 (<m_{j; j\neq i, , m_j>m_i}> -m_i), \qquad (24)$$

where the coefficient 0.6 is obtained for n=4, $r_p$=3, and $r_1$=1.6. The term $<m_{j; j\neq i, , m_j>m_i}>$ denotes the mean cooperation level of the n-1 other group members, where the average is conditioned on the contributions being larger than $m_i$.

**5.2 Comparison between the theoretical punishment level and experimental results**

Expression (24) predicts the expenditure by punishing group members as a function of the deviation from the mean cooperation level of the other group members. Figure 4 compares Fehr and Gächter (2002)'s experimental results shown in their figure 1 with the theoretical expenditure by punishing group members predicted by expression (24). For the experimental data, we use the average of the mean expenditure over all the six periods studied by Fehr and Gächter (2002). We augment our prediction (24) by providing a standard deviation obtained as follows. We simulated 20'000 synthetic random games with $r_p$=3, and $r_1$=1.6 and n=4 players. Each agent has the same punishment coefficient k given by (23) and her contribution $m_i$ is taken from a uniform distribution between 0 and 20. The punishment she exerts on other agents is given by the rule (18).

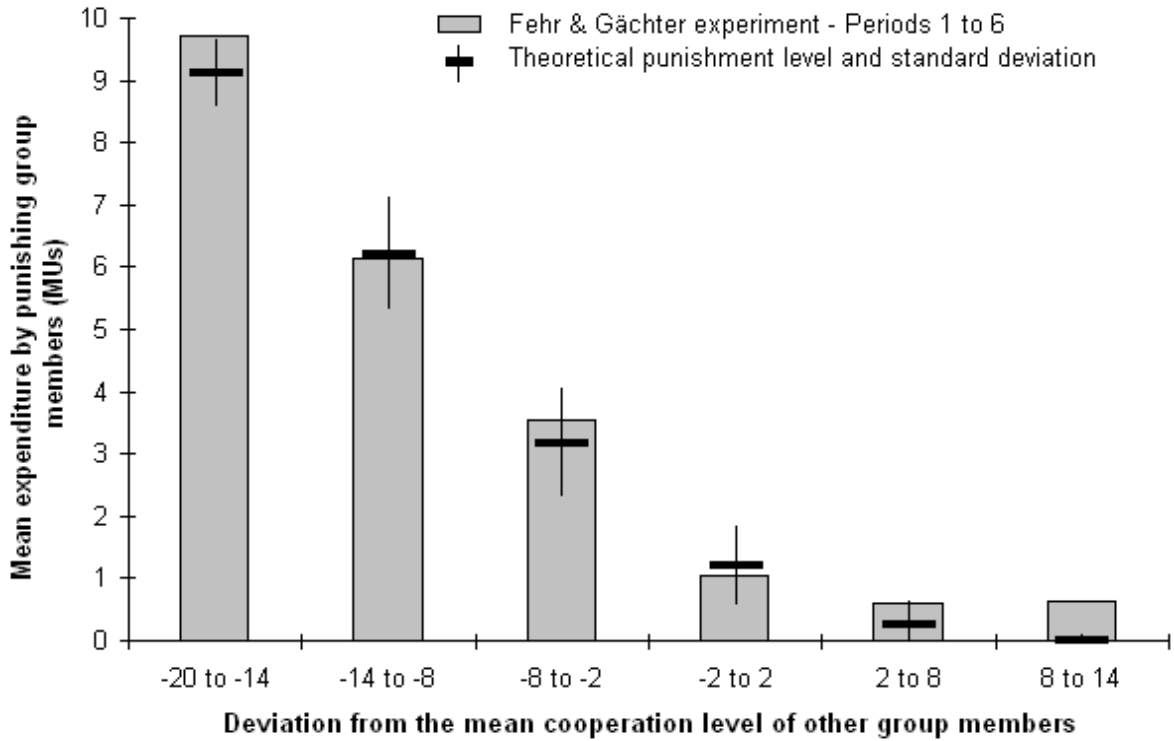

**Fig.4:** Mean expenditure by punishing group members (in MUs) as a function of the deviation between the contribution of an agent from the mean cooperation level of the other group members. The vertical bars are the averages of the mean expenditure over all the six periods studied by Fehr and Gächter (2002). The thick horizontal segments give the punishment level predicted by expression (24). They are accompanied by vertical segments indicating ± 1 std, obtained as described in the text.

The agreement between prediction (24) and Fehr and Gächter (2002)'s experimental results is striking, especially given the fact that there are no adjustable parameters. The only discrepancy occurs for the largest *positive* deviation 8-14 from the mean cooperation level of other group members. For strictly positive deviations, rule (18) implies no punishment. The fact that we predict some punishment for a positive deviation in the range 2-8 results from the heterogeneity of the agents' contributions: even if the deviation of agent i's contribution from the mean contribution is positive (namely, agent i gives more that the average), there is the possibility that some other agent j has contributed still even more (while at least one of the two others has necessarily contributed less), in which case agent i is susceptible to agent j's punishment, according to (18). For the largest *positive* deviation 8-14 from the mean cooperation level of the other group members, this occurrence is not observed in our simulations as it becomes essentially impossible to have an agent contributing much more that the average of the others, and still find one of them contributing even more. The discrepancy between our prediction (24) and the data for the deviation 8-14 thus results from the punishment by low contributors of large contributors, an effect not taken into account in our theory.

**6-A simple model of repeated agent interactions with punishment**

Our interpretation and use of the evolutionary feedback selection theory states that the average level - and probably the full distribution as well - of feedbacks by reward and/or punishment has been tuned over many generations of human beings by repeated interactions in which humans strive to optimize their own selfish gains in the presence of the other agents' feedbacks. Using this concept, we have been able to account quantitatively for several experiments on social cooperative dilemma games. Now, we present a simple model of evolution of players, in order to test whether the level of punishment indeed evolves or not to the optimal equilibrium predicted by our theory of equilibrium (see expressions (8), (12) and (23) for three cooperative games with punishment). In other words, our goal is to check whether or not the level of punishment predicted by the equilibrium argument applied above is indeed selected by some reasonable implementations of the dynamics of interacting agents. Our simulations also address the question of whether a population of agents with optimal punishment level remains stable against random variations and inclusion of invaders with different punishment levels.

We consider a synthetic universe of n agents who play together repeatedly the game of section 5, over T=500 periods. At a given time t, each agent i cooperates according to her frozen contribution $m_i$ and punishes other agents according to rule (18) with a personal frozen punishment coefficient $k_i$. Each agent is thus characterized by a fixed pair of parameters ($m_i$, $k_i$), where $m_i$ is uniformly drawn between 0 and 20 and $k_i$ is uniformly drawn between 0 and 1. After each game session, each agent sees her "wealth" change according to her profit-and-loss, which includes the revenue from the group project and the losses from her punishments to others and from others to herself, as well as a fixed consumption equal to 3 MU per session. "Wealth" is like a monetary reserve for food and other exchanges of life-supporting commodities. The cooperative group project mimics a cooperative hunter-gatherer food collection. The consumption is introduced to obtain a stationary wealth and allows implementing a selection process in which big losers disappear. The initial "wealth'' is set to 50 MUs for each agent.

The updates, which allow for both adaptation and evolution, are as follows.
(i)     Adaptation: if the profit-and-loss of agent i at time t is negative, she modifies her ($m_i$, $k_i$) by small random unbiased increments around her previous values. This step represents the natural ability of agents to explore other solutions when failing in the recent past. Here, we call `adaptation' the generation of random variations within an agent in response to failure, such as for instance the proposition that the genome may be lowering the activity of its own repair mechanisms in response to environmental stress. We have found this ingredient to be secondary for the determination of the equilibrium, while it controls significantly the duration of the transient dynamics.

(ii) Evolution: if the wealth $W_i$ of agent i becomes smaller than $m_i$, she disappears and is replaced by another agent endowed with a punishment coefficient $k_j$ and a contribution $m_j$ drawn from uniform distributions centered on the respective averages all the other agents of the universe. This step reflects the genetic mixing among humans along the course of their co-evolution and the cultural influence of group members towards new generations of participants. Here, the variations are between individuals rather than within individuals in (i).

(iii) The global wealth of all agents has no real significance and may increase (inflation) or decrease (deflation). To reflect the invariance of the cost of consumption with respect to a global change of wealth, we change the consumption cost to ensure an average stationary global wealth: if the total wealth increases (resp. decreases) above (resp. below) the initial wealth (50n), the consumption cost per agent and per unit time is increased (resp. decreased) by 0.5 MU for the next time step.

We stress that ingredients (i) and (ii) refer respectively to variations within and between individuals, both being subjected to selection and thus result in evolved adaptation (natural selection on variations). Moreover, this agent-based model does not incorporate the third level of variation, in terms of group selection. This does not imply that we take this third level of variation as negligible. Our goal here is simply to demonstrate, with perhaps the simplest possible model, the plausibility of our hypothesis that evolutionary feedback selection will determine optimal levels of cooperation that can be predicted from a self-centered cost-benefit analysis taking into account feedbacks from the group.

We have run this game 200 times for 500 time steps for various values of the parameters $r_l<n$, $r_p>1$ and $n>1$, and have constructed the distribution of punishment coefficients k of the evolved population of agents in the last 200 time steps to remove possible biases from initial transients. For $r_p=3.5$, we find that the distribution of k's develops a mode centered very close to the prediction (23) derived from our cost-benefit use of the evolutionary feedback selection theory. Fig 5 shows that prediction (23) is able to account quantitatively for the non-monotonic dependence of k as a function of the number n of agents in a group.

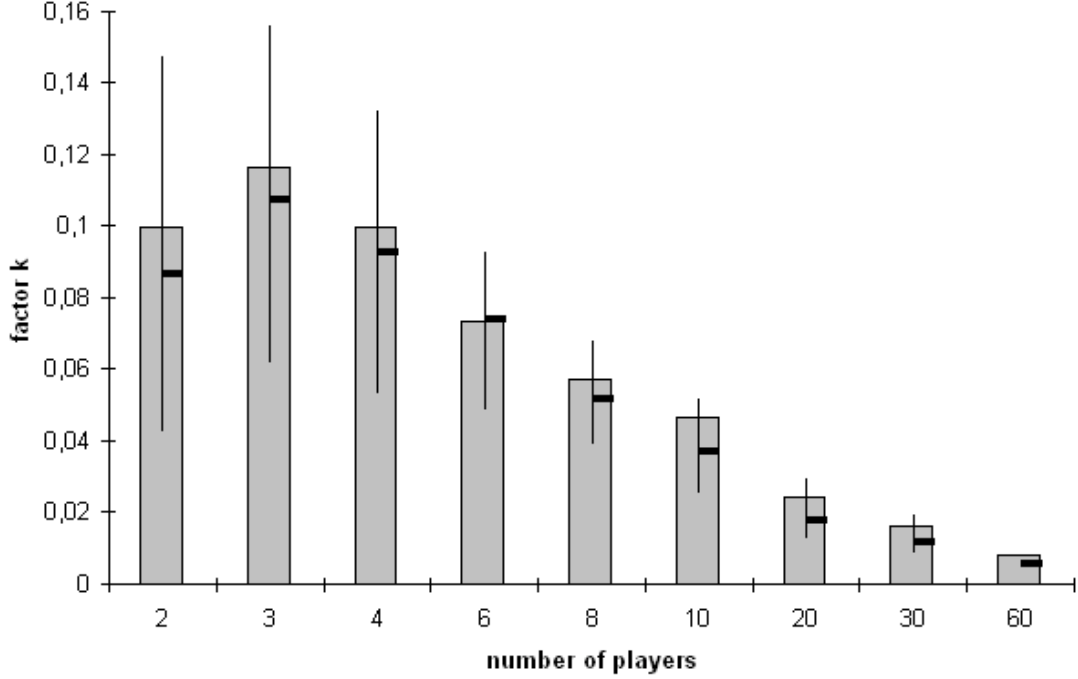

**Fig.5:** Dependence of the $k*r_1$ - the linear punishment response function factor time $r_1$- as a function of the number n of agents in a cooperating group. The vertical bars are the predictions of the theory given by (23), while the vertical segments define the 25% quartile (lower endpoint), the median (thick middle horizontal segment) and the 75% quartile (upper endpoint) obtained from simulations of the agent-based model with parameters $r_1=1.6$ and $r_p=3.5$. Note the nonlinear scale used in the abscissa.

Our simulations also find an elaborate balance between diversity, punishment and cooperation, controlled by the punishment efficiency $r_p$.

(1) For small $r_p$'s, punishment which promotes cooperation would require a large propensity k to punish which should be selected simultaneously by a large fraction of the population, which is unlikely. As a consequence, strong punishers (i.e., with large k's) find themselves rather isolated and have a high probability to be eliminated by the selection process due to the cost paid for punishment. Then, cooperation rapidly vanishes and the average contribution $<m_i>$ is close to zero. No cooperation (with everyone contributing little or nothing) implies low or no global punishment since the diversity of contributions is low or zero. Consequently, the diversity in the distribution of k's becomes irrelevant as the act of punishing disappears with the absence of cooperation. In this non-cooperative equilibrium, there is no co-evolution of cooperation and feedback by punishment.

(2) In contrast, large $r_p$'s increase the probability of reaching a stable cooperation regime with minimal costs incurred by punishers during transient regimes. Indeed, an agent contributing more than the average and with a strong propensity to punish will be quite effective alone in promoting cooperation at a tolerable cost.

(3) In between, one can expect a transition between low $r_p$ and large $r_p$'s in which the global cooperation level and the average propensity to punish are held in balance by a sustained stable diversity of cooperation.

Figure 6 illustrates these three regimes by plotting as a function of $r_p$ the mean contribution $<m_i>$. The large range of mean contribution levels in the regime (3) of intermediate $r_p$'s reflects the sensitive dependence of cooperation from one realization to another realization. Equations (21) and (22) indicate that the optimal level for the propensity k to punish depends on the final distribution of contributions, measured by $F(<m_i>)$ (which is defined as the probability that any participant invests less than or equal to the average). In our simulations, we observe that $F(<m_i>)$ remains close to ½ from $r_p=1$ up to $r_p=4.5$ and then grows while exhibiting large fluctuations from one realization to another. As a consequence, the sensitivity of the factor k to $r_p$ is found smaller than anticipated by equation (23), where $F(<m_i>)$ is approximated by ½ for all regimes.

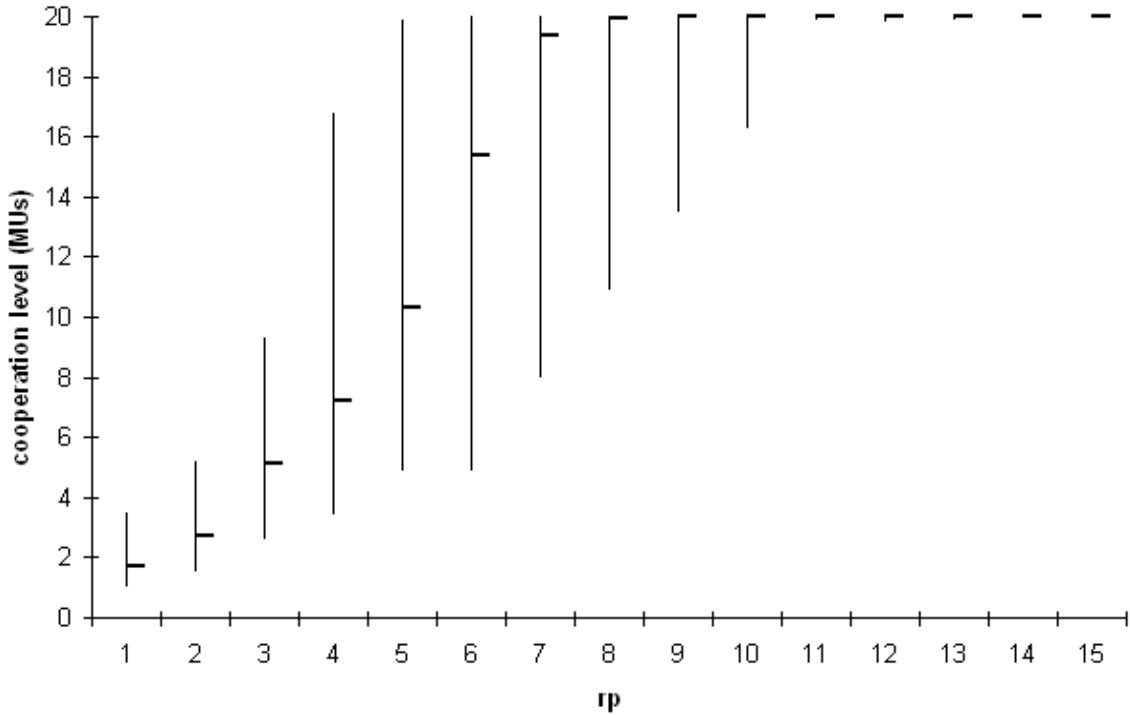

**Fig.6:** Dependence of the cooperation level – measured by the 25% quartile (lower endpoint), the median (thick middle horizontal segment) and the 75% quartile (upper endpoint) obtained from simulations of the agent-based model – as a function of the punishment efficiency $r_p$, for n=8 and $r_1$=1.6.

While this simple agent-based model does not claim to represent faithfully the evolutionary process and co-evolution of genes and culture of human beings (Richerson and Boyd, 2005), it suggests that the level of feedback, quantified by the punishment coefficient k, is an emergent property of the co-evolution of interaction agents with feedbacks. In other words, the level of punishment that evolutionary feedback selection can be seen as an attractive stable point of the dynamics of human interactions.

The reader may wonder what are the profound relationships between such a simple agent-based model and the theory of evolutionary feedback selection that we have quantitatively exploited to predict the average propensity to punish in three different games. Indeed, the theory of evolutionary feedback selection postulates that agents have endogenized the expected behavior of others in their own mind, and it is this capacity that allows them to tune their propensity to punish to an optimal equilibrium level in a cost-benefit sense. Our agents in the numerical simulations do not have such sophisticated brains! They are either adapted by luck and thrive, or they die and other descendants replace them, pushing the group to evolve towards a dynamical equilibrium as predicted by the theory of evolutionary feedback selection. This is the key point that we wish to emphasize: our agents in the model simulations actually collectively solve the global evolutionary optimization problem. However, we should stress that there is a large difference between humans and the artificial agents in our simulations. The former have adapted and endogenized pro-social preferences and social norms that they can use in novel contexts. In contrast, our artificial agents are adapted to only one universe, the restricted artificial world of the repeated altruistic punishment game. In other words, the synthetic agents in our present simulations have not developed the pro-social abilities that would allow them to adapt and generalize their behavior to novel situations. Many other limitations characterize our present simulations which will be addressed in the near future (Hetzer and Sornette, in preparation).

**7-Reward versus punishment**

The emphasis of our analysis has been on the experiments in which feedback may occur via punishment. But the evolutionary feedback selection theory is not one of feedback by punishment alone. It rests on the general notion that feedbacks of any kind may participate in favoring long-term cooperation between humans. One such feedback is reward, which is technically symmetric to punishment: when an agent has done better than expected according to some reference, other agents may reward her. This leads us to suggest variations of the games analyzed here and others (in addition to the gift exchange games (Fehr et al., 1993), trust games (Berg et al., 1995) or sequentially played prisoners' dilemmas (Hayashi et al., 1999), in which punishment is replaced by reward.

We have argued above for the co-evolution of cooperation and feedback by punishment, probably aided by organization of psychological and emotional brain modulii. Assuming a similar co-evolution of cooperation and feedback by reward, we can apply the cost-benefit analysis of the evolutionary feedback selection theory to the game of section 5 in which punishments are replaced by rewards. Assuming that the reward $Re_{j \to i}$ of an agent j to another agent i follows a rule similar to (18) (namely $Re_{j \to i} = k_r (m_j - m_i)$, for $m_i > m_j$; $Re_{j \to i} = 0$, for $m_i \leq m_j$), and using a formalism parallel to that of section 5, we find that expression (21) is changed by replacing $1-F_i$ by $F_i$ and vice-versa, while expression (23) still holds for the coefficient of reward $k_r$ but with $r_p$ now interpreted as the efficiency of a reward. Similar results are obtained for the third party game of section 3 in which rewards replace punishments. Should one believe these predictions?

Feedback by punishment is probably associated with "negative" emotions, such as anxiety, anger, fear, shame and guilt, at various degrees either for the punisher and the punished one. In contrast, feedback by reward may trigger different kinds of emotions, such as desire, hope, joy, and pleasure (Harbaugh et al., 2007). Recent advances in evolutionary psychology and neurobiology (Damasio, 1994; 1999) indicate that different emotions are associated with distinct complex activations of parts distributed over many locations in the brain. In addition, the efficacy and robustness of punishment as a feedback mechanism seem much larger than reward, as interestingly captured by Niccolo Machiavelli in The Prince: "… Upon this a question arises: whether it be better to be loved than feared or feared than loved? … because it is difficult to unite them in one person, it is much safer to be feared than loved … for love is preserved by the link of obligation which … is broken at every opportunity for their advantage; but fear preserves you by a dread of punishment which never fails."

**8-Concluding remarks**

According to the theory of "strong reciprocity" (Gintis et al., 2005), humans care about fairness and the welfare of others beyond what can be explained by evolutionary kin theory, costly signaling theory and indirect reciprocity, as well as reciprocal altruism or direct reciprocity. Humans are characterized by extraordinary cooperation among themselves, with strong feedbacks through reward and punishment. Natural and cultural selections are arguably the most promising mechanisms to explain these prosocial behaviors (Henrich, 2004). We have referred to these mechanisms taken collectively as "evolutionary feedback selection" in order to avoid the on-going discussions concerning the relative importance of cultural versus natural selections and the roles of within-group and between-groups selection. We have proposed that evolutionary feedback selection implies an equilibrium (or dynamical self-organized fixed point) characterizing the propensity to feedback (analyzed here for punishments), which can be determined quantitatively by a simple self-centered cost-benefit analysis. Our approach emphasizes the role of feedbacks, without specifying if it occurs by natural and/or cultural evolutionary selections, which lead to the selection of behaviors well described by strong reciprocity. These feedbacks may occur at, as well as promote, different levels of selection (within or between individuals and groups). Notwithstanding the fact that our arguments have been framed in terms of individual incentives leading to individual selection along the line of Tooby and Cosmides (1996), with a derived role for group selection, our formalism could as well apply to the group level.

According to the theory of strong reciprocity, people maximize a utility function, which is the sum of a selfish gain and of a term favoring fairness (Fehr and Schmidt, 1999). Evolutionary feedback selection focuses on the evolutionary origin of these conflicting preferences by emphasizing that people have evolved to maximize their selfish gain in the presence of feedbacks. Note that feedback constraints can always be endogenized in a Lyapunov function using Lagrange multipliers (Hahn, 1963), leading to a reinterpretation in terms of a utility function. By stressing the importance of feedbacks in an evolutionary dynamical framework, the evolutionary feedback selection provides a sound underpinning and understanding of human cooperation, as witnessed by its ability to provide quantitative predictions to experiments. Evolutionary feedback selection differs from group selection theory (Wilson and Sober, 1994), as it does not necessarily rely on the competition between and/or the selection of groups.

While the extended utility approach to strong reciprocity implies that punishment is performed as a tool to recover fairness, evolutionary feedback selection emphasizes a different driving force: punishment is not performed to recover fairness but as a tool promoting the maximum selfish gain in the presence of possible cooperation which co-evolved with it. This interpretation is indeed supported by experiments, which show that fairness is not recovered nor obtained by the action of punishment.

We have illustrated how the general theory of evolutionary feedback selection can be implemented through a simple self-centered cost-benefit analysis, by taking examples in which the feedbacks occur via punishments. These examples are a third party punishment game, the ultimatum game and an altruistic punishment game. Satisfactory quantitative agreement between the predictions on the propensity to punish and its empirical measurements in these three games has been obtained. The general conceptual conclusion that can be drawn from our analysis is that what is called altruism and selfishness might not be paradoxical behaviors. They are in fact the two faces of Janus. Looking back in the past, this suggests that the Darwinian selection process has promoted a biological and cultural reconciliation between self-regarding and others-regarding preferences. Feedback by punishment (and probably also by reward and other mechanisms) might have been the trigger for adjusting social emotions to a level ensuring group efficiency, cascading down to individual self-interest, and reciprocally cascading up to the group level. Evolutionary feedback selection addresses human rationality in a systemic and evolutionary manner. An individual's rationality, her social emotions, her ability to understand others' desires, to transmit knowledge and social learning, influences group efficiency on a longer time scale. Survivors are rationally efficient from an ecological perspective. The extensive cooperation within human societies, the biological development of a human "social brain" (Dunbar, 1998) and the adjustment of reward-punishment to a fine-tuned level constitute co-evolutionary processes, strengthening human groups in a Darwinian selection process. Evolutionary feedback selection as we understand it in terms of a self-organizing attracting dynamical equilibrium bypasses by construction the second-order free-rider problem (Panchanathan and Boyd, 2004).

We have discussed situations where feedback by punishment happens, which promotes cooperation. A natural extension is to ask how some cooperative behavior can occur in initial rounds and in non-repeated games in the absence of punishment or other feedbacks, as for instance reported in (Fehr and Gächter, 2002; Fehr et al., 2002). The theory of evolutionary feedback selection provides a natural explanation of this often reported behavior: initial rounds should express the prior estimation by subjects of what is the expected relevant social normal to activate in the novel situation. This prior should result from the cumulative impact of the previous "games" experienced by the subjects on cooperation and its absence, in the presence of all the different types of feedbacks that may appear during a lifetime. This is this reasoning which is at the origin of the choice to replace $F_i$ defined in (22) by its expectation $E[F_i]=1/2$ over all possible histories (that is, over all possible distributions of others contributions) in the case of the altruistic punishment game. Since a typical human being experiences from his very young age many situations where cooperation is called for, and is promoted by all kinds of feedbacks, it is natural to expect a spontaneous "firing" of the average social norm that the subject has endogenized. Now, if such a game is repeated without the existence of feedbacks, cooperation is known to decrease and vanish eventually (Fehr and Gächter, 2002; Fehr et al., 2002). The explanation for this observation is that the subjects learn to optimize and find the corresponding Nash equilibrium (Binmore, 2007). Only the reactivation of punishment can bring back cooperation (Fehr and Gächter, 2002; Fehr et al., 2002).

Looking forward, competition could keep on favoring larger groups of connected individuals, sharing common views and reciprocal fairness, enforced by Education and Law. The future could however look very different. The optimization process that we outlined regarding altruistic emotions is not unconditional. It constitutes an ongoing process of local selection under ecological constraints. Our cost-benefit equilibrium analysis does not claim that altruism is "optimal" but only that it has evolved to adapt to some situations that have been experienced by our ancestors over many generations, some of these situations being partially proxied by the three games analyzed he re.

**Acknowledgements:** The ideas presented here have benefited from stimulating exchanges with Rob Boyd, Robin Dunbar, Alan Fiske, Stephen Le, Karthik Panchanathan, Jeff Satinover and the Human Complex Systems group at UCLA. We are grateful to Riley Crane, Urs Fischbacher, Georges Harras, Moritz Hetzer, Stephen Le, Yannick Malevergne and Francis Steen for helpful feedbacks on the manuscript.